\documentclass[12pt,preprint]{aastex}

\def\halpha{H$\alpha$}

\def\gradF{$\nabla F$}
\def\barF{$\bar F$}
\def\matrixA{{\bf A}}
\def\ergscmsec{${\rm ergs~cm^{-2}~sec^{-1}}$}
\def\la{\lessapprox }
\def\ga{\gtrapprox }

\begin{document}


\title{A Robotic Wide-Angle H$\alpha$ Survey of the Southern Sky}

\author {John E. Gaustad}
\affil{ Dept. of Phys. \& Astr., Swarthmore Coll.,
Swarthmore PA 19081, USA }
\email{jgausta1@swarthmore.edu}

\author{Peter R. McCullough\footnote{Cottrell Scholar of Research Corporation}}
\affil{ Dept. of Astronomy, Univ. of Illinois, Urbana IL 61801, USA }
\email{pmcc@astro.uiuc.edu}

\author {Wayne Rosing}
\affil{ Las Cumbres Obs.,
1500 Miramar Beach, Montecito CA 93108, USA}
\email{wrosing@lco.org}

\and

\author{Dave Van Buren}
\affil{ Extrasolar Research Corporation,
Niskayuna NY 12309, USA}
\email{vanburen@extrasolar.com}


\begin{abstract}
We have completed a robotic wide-angle imaging survey of the
southern sky ($\delta = +15\degr $ to $-90\degr $) at 656.3 nm wavelength, 
the H$\alpha$ emission line
of hydrogen. Each image of the resulting Southern H-Alpha Sky Survey Atlas 
(SHASSA)
 covers an area of the sky
$13\degr $ square at an angular resolution of approximately 0.8
arcminute, and reaches a sensitivity level of 2 rayleigh 
($\rm 1.2 \times  10^{-17} ~erg~cm^{-2}~s^{-1}~arcsec^{-2} $) per pixel,
corresponding to an emission measure of $\rm 4~cm^{-6}~pc$, and to
a brightness temperature for microwave free-free emission of 
$\rm 12 ~\mu K$ at $\rm 30 ~GHz$. Smoothing over several pixels allows
features as faint as 0.5 rayleigh to be detected.
\end{abstract}



\section{Introduction and Scientific Purpose}  \label{intro}

Early surveys of the Milky Way, such as those of \citet[]{bar27}, photographed 
several regions of interstellar gas ionized by the ultraviolet radiation of 
nearby hot stars. Many other surveys were done over the ensuing decades (see the 
list of references given by  \citet[]{siv74}). However, it was not until after 
the discovery of pulsars  \citep[] {hew68} that it became generally recognized 
that ionized hydrogen is a pervasive component of the interstellar medium. Soon 
thereafter  \citet[]{siv74} conducted a photographic survey in search of this 
general diffuse medium, reaching a limiting brightness of 15 rayleigh 
($\rm 1~R = 10^6/4\pi ~photons~cm^{-2}~s^{-1}\Leftrightarrow   EM \approx   
2~cm^{-6}~pc$), corresponding to an emission measure of about $\rm 30  ~cm^{-
6}~pc$, and covering the band of latitude $\rm |b| <  25\degr $ at a resolution 
of approximately 10$\arcmin $. Somewhat later, \citet[]{pgk79}, using an image 
tube camera, published an atlas of images in four emission lines ([Sll], 
H$\alpha $+[NII], O[III], H$\beta $) and the blue continuum (422 nm), covering 
the region $\rm |b| < 8\degr $, reaching a surface brightness ($\approx 15$ R) 
about the same as that of Sivan's survey, at an angular resolution of $0\farcm6 
$. Of course the H$\alpha $ line is included in the pass band for the red plates 
of the National Geographic Society/Palomar Observatory Sky Survey (POSS 1), and 
these show many prominent emission nebulae, those brighter than approximately 
100 R. (The ESO/SRC Southern Sky Atlas is similar, but goes a bit deeper.) The 
second generation plates (POSS 2) go about a factor of four fainter, reaching a 
level of approximately 25 R. 

Two difficulties with these photographic surveys are that their sensitivity is 
low, and because of the non-linear response of photographic emulsions and 
contributions from continuum radiation it is difficult to derive quantitative 
intensity information from them. Modern CCD detectors, on the other hand, are 
intrinsically linear,  have high quantum efficiency, and allow for digital 
subtraction of continuum images. Although photographic surveys still have their 
place \citep[]{par98}, most modern studies are done via narrow-band CCD imaging 
or Fabry-Perot spectroscopy. An early example of the former technique is the 
work of \citet[] {mrt90}, who surveyed the region $l = 120\degr $ to $190\degr 
$, $ b = -30\degr $ to $+15\degr $ to a level of 4 R ($\rm 8~cm^{-6}~pc$) at 2 
arcminute resolution.  An early example of the Fabry-Perot technique is the work 
of \citet[]{rey73} who observed several regions with a Fabry-Perot spectrometer. 
Reynolds and his collaborators have recently completed a survey of the northern 
sky with a Fabry-Perot instrument  called the Wisconsin H-Alpha Mapper (WHAM) 
\citep[]{rey98, haff01a, haff01b}. A group at Virginia Polytechnic Institute 
\citep[]{dst98} have begun a narrow-band  survey of the northern hemisphere with 
an imaging camera similar to ours. We present a  comparison of these other 
surveys with our own in Section \ref{others}.

Both of these narrow-band imaging surveys are complementary to Reynolds' Fabry-
Perot survey, not competitive with it. Our angular resolution is 75 times better 
than that of WHAM. The WHAM spectrometer, on the other hand,  goes ten times 
fainter. It also has high spectral resolution (0.026 nm, $\rm 12 ~km~s^{-1} $), 
and thus provides  velocity as well as  intensity information  that the narrow-
band imaging surveys cannot. Perhaps as important, the high spectral resolution 
allows the WHAM instrument to separate interstellar H$\alpha $ emission from 
that coming from the Earth's geocorona and thus determine  absolute intensities, 
whereas the images provided by the narrow-band surveys all contain an offset due 
to geocoronal emission (as well as telluric OH emission), an offset which is 
variable both temporally and spatially. In Section 5 we describe a technique for 
correcting our images for geocoronal emission that uses WHAM observations as 
fiducial points in those parts of the sky where our surveys overlap.

We present this survey to the astronomical community as a resource which we hope 
will find many applications. We anticipate its use in two major areas of study: 
first, the structure of the interstellar medium, and second, the 
characterization of the Galactic microwave foreground.

The survey will provide detailed information on the
structure of the diffuse, warm, ionized component of the
interstellar medium, information necessary for understanding the
dynamics and evolutionary history of the interstellar gas. Since
the survey  covers the entire  sky south of $+15\degr  $
declination, has higher angular resolution and reaches a fainter intensity level 
than does any
previous large scale survey, we expect a detailed examination will show several 
new structures,
such as stellar wind bubbles, supernova remnants, bow shock
nebulae, planetary nebula shells, and other interesting
objects. Some examples are noted in Section \ref{examples}.

Even where  specific H$\alpha$ emission features are not seen, the survey images
should be scientifically useful. Since the production of both kinds of
emission results from the interaction of a proton with an
electron, the ratio of intensities of 
free-free and  H$\alpha$ emission  depends primarily on
quantum-mechanical factors, with only a slight dependence on
temperature and no dependence on density \citep[] {vgb98}:
\begin{equation}
{{T_b^{ff} [mK]}\over{I_{H \alpha } [R]}} = 10.4 \nu^{-2.14} T_4^{0.527} 
10^{0.029/T_4} (1 + 0.08)
\end{equation}
where $\nu$ is the observing frequency in GHz, $T_4$ is the temperature of the 
gas in units of $10^4$ K, and
the second term in the last factor (1+0.08) is the relative abundance of helium.
Thus our H$\alpha $ images will
allow accurate measurement of anisotropies in the Galactic
free-free emission at microwave wavelengths (or proof that these
are negligible), emission which must be subtracted from satellite
or ground-based measurements to obtain the spatial spectrum of the true cosmic 
background
radiation. In an early pilot study for this
project \citep[] {gmv96} we showed that no more than 7\% of
the 44 $\mu K$ anisotropy in the CMBR measured in the Saskatoon
experiment \citep[] {net95} could be due to contamination
by Galactic free-free emission. The high angular resolution of our survey may be 
important given the discovery of a third acoustic peak in the microwave 
background spectrum at angular scales of about $0\fdg2 $ \citep[] {boom01}, well 
within the range of spatial frequencies we sample. A general review of the 
implications of H$\alpha $ observations for studies of the cosmic microwave 
background can be found in \citet[] {mgrv99}.

\section{Instrumentation and Observational Procedure} \label{obs}

\subsection{Camera, Telescope Mount, and Dome} \label{equipment}

Our CCD camera, from SpectraSource, Inc., contains a 1024 x 1024 Texas 
Instruments chip with the following characteristics:  12-$\micron $ pixels, 
full-well capacity  80 Ke,  quantum efficiency  62\%,  gain  1.4 e ADU$^{-1}$, 
and readout noise  18 e. It is cooled thermoelectrically to about $-35\degr $C, 
yielding a dark current of  0.04 e s$^{-1}$. In a 20-minute exposure, the 
maximum we use, the dark level amounts to  about 34 ADU (48 e), about 10\% the 
level of our typical exposures. Laboratory measurements showed that the chip 
response was linear up to the full-well capacity  (\rm 80,000 e $\approx $ 
60,000 ADU), but the variance increased non-linearly above 20,000 ADU (28,000 
e). Except for some very bright objects, our exposures yield values well below 
this limit.

The sky is imaged onto the chip with a 52-mm focal length Canon lens operated at 
f/1.6, yielding a field of view of $13\arcdeg \times 13\arcdeg  $ and a scale of 
$47\farcs 64$ pixel$^{-1}$.

The camera is attached to a Byers Series 2 German mount with microstepping 
motors  on each axis.  The pointing  accuracy is better than $5\arcsec $ up to a 
zenith distance of 75\degr . The unguided tracking accuracy is better than 
$2\arcsec $ hr$^{-1}$. Thus, with $48\arcsec $ pixels, no external guider is 
needed.

The system is housed in a 10-foot dome (from Technical Innovations, Inc.) on a 
concrete pad located near the 1-meter telescope building at Cerro Tololo Inter-
American Observatory (CTIO). The small size of the instrument and dome has 
caused the CTIO staff to give our instrument the nickname of El Enano---The 
Dwarf. We added magnetic microswitch sensors, power relays, and a position 
encoder to the dome system so that it could be controlled by a computer. We also 
added a third dome rotation motor, replaced some plastic parts with metal, and 
modifed the shutter cable system in order to improve reliability of the dome 
operation.  A Texas Weather Instruments weather station completes the system. 
Two PCs running scheduling and control programs written in Visual Basic under 
Windows 95 control the camera, mount, dome, weather station, and light box. They 
communicate with each other and with the outside world via Ethernet using 
standard TCP/IP protocol.

We are grateful to  Las Cumbres Observatory, Inc., for loaning us much of this 
equipment for the duration of the project.

Because the twilight sky is not uniform over our wide field of view 
\citep{chr96}, we do not use sky flats for calibration. Instead we use a 
lightbox  containing a translucent screen illuminated by incandescent light 
bulbs and by hydrogen emission lamps.

\subsection{Filters} \label{filters}

A filter wheel mounted between the lens and the sky contains an H$\alpha$ filter 
of 3.2-nm bandwidth and a dual-band notch filter, which excludes H$\alpha$ but 
transmits two 6.1-nm bands of continuum radiation on either side of H$\alpha$, 
centered at 644 nm and 677 nm (Figure \ref{FilterTrans}). Both filters were 
obtained from Custom Scientific, Inc. 

The choice of the H$\alpha $ filter bandwidth was governed by two 
considerations. We want a narrow bandpass to reduce  the foreground sky 
continuum, but  large enough  that transmission losses near the edge of the 
field are small.  Because our filter is placed in front of the camera lens, 
light imaged at any point on the detector enters the filter as a parallel beam 
at an angle of incidence equal to the angular distance of the point from the 
optical axis. For a simple interference filter (multi-layer filters are somewhat 
more complicated) the central wavelength at angle of incidence $\theta $ is 
given by the formula \citep{smith90}
\begin{equation}
\lambda = \lambda_{o} \sqrt {{1 - {{\sin ^2 \theta } \over {n^2 }}}}
\label{cwlshift}
\end{equation}
where $n$ is the index of refraction. The filter band shifts towards the blue 
with increasing angle of incidence; equivalently, a line shifts towards the red 
with respect to the filter central wavelength, and moves to a different part of 
the filter transmission curve.  For $\theta = 9 \arcdeg  $, near the corners of 
our images, the wavelength shift at 656.3 nm is approximately 1.0 nm. We have 
therefore chosen a filter with 1.6 nm half-width.

The central wavelength of an interference filter  shifts to longer wavelengths 
with increasing temperature, in our case about 0.03 nm \arcdeg C$^{-1}$. The 
filter wheel is encased in styrofoam and maintained at a constant 25 \arcdeg  C 
temperature to minimize such temperature-related shifts and to inhibit moisture 
from condensing on the filters or the lens. The temperature is deliberately set 
higher than the design temperature of 20 \arcdeg  C, and higher than the night-
time ambient temperature, so that
a) only active heating, not cooling, is needed to maintain the temperature;  
b) the bandpass is shifted slightly to the red of H$\alpha $ at the center of 
the field, thus reducing the variation in transmission with angle of incidence 
as the central wavelength shifts blueward towards the edge of the field; and c) 
the possibility of condensation on the lens is eliminated. 
The transmission of 656.3 nm light as a function of angle of incidence, as 
measured on our lightbox while illuminated by a hydrogen emission lamp, is shown 
in Figure \ref{HFilterTrans}.

Figure \ref{FilterTrans} also shows the position of atmospheric OH emission 
lines and interstellar emission lines.
The H$\alpha $ band (653.3 - 659.3 nm for T/T$_{max} > 10\% $ ) contains only 
five lines of the 6-1 band of OH. But some of these, particularly the 655.4 nm 
line, may make a significant contribution to the general foreground level, 
comparable to that from geocoronal H$\alpha $. To the extent that this 
foreground emission can be modelled as a 2-dimensional plane, it is removed from 
our final images by the methods discussed in Sections \ref{corrections} and 
\ref{georemoval}.
In addition, the interstellar emission lines of [NII] at 654.8 and 658.3 nm lie 
within the H$\alpha $ filter band. The filter transmission is approximately  
39\% and 26\% respectively at these wavelengths (at the operating temperature of 
25 \arcdeg  C) compared to the transmission at H$\alpha $ of 78\%. In some 
astronomical objects, the [NII] lines are comparable in strength to H$\alpha $, 
so this contamination by [NII] lines should be taken into account in any careful 
quantitative measurements of the atlas images. For example, if the 658.3 line 
has the same intensity as H$\alpha$ and the 654.8 nm line has 1/3 that strength 
(the ratio of the two [NII] lines is always 1/3), about 33\% of the brightness 
measured in the H$\alpha $ pass band is really [NII] emission, not H$\alpha $. 
(See also the discussion of the [NII] lines in Section \ref{intensityscale}.)

The continuum filter bands were chosen primarily to avoid  emission lines in the 
night sky \citep[]{brkn68, ostr92,ostr96}. The 644 band falls longward of the 
stronger lines of the 9-3 band of OH and excludes most of the short wavelength 
lines of the 6-1 band of OH, as well as excluding the OI lines at 630.0 and 
636.4 nm. The 677 band falls between the 6-1 and 7-2 bands of OH, and excludes 
the interstellar lines of HeI  at 667.8 nm and of [SII] at 671.6 nm. The 
transmission at the wavelength of the [SII] line at 673.1 nm is only 16\%, 
though it will be somewhat higher, up to 70\% , at high angles of incidence. In 
many astronomical objects this line is less than 10\% the strength of H$\alpha 
$, though it can be of comparable strength in shocked regions, such as in 
supernova remnants, and in the very faint emission from the warm interstellar 
medium \citep[]{hrt99}. The exposure times used for the continuum exposures are 
four times shorter than for the H$\alpha $ images, so the contamination from the 
[SII] 671.6-nm line is probably no more than a few per cent in most regions, 
though it may be as high as 25\% in strongly shocked regions which appear near 
the edges of our fields.

\subsection{Computer Control} \label{computercontrol}

One of the robot's computers positions the telescope and dome while the other 
operates the camera, schedules the observations, and runs the ftp and email 
communication links. A log of the night's operations and sample images are sent 
to the PI each morning via email and ftp, but the full data set is recorded on 
disk and then transferred to DAT tape during the day. We can inquire of the 
status of the robot and modify its observing schedule via email, and make 
changes in the control programs via ftp. But we do not operate the camera in 
real time--it is a true robot, in the sense that its real-time operations are 
autonomous . The robot monitors the weather station continuously, and will not 
attempt observations if humidity or wind conditions exceed safe limits, but as 
an extra precaution it also asks permission (via email) of the 4-meter telescope 
operator before opening the dome. The operator can also close the dome via email 
command. The only other human intervention, during normal operations, is for 
weekly tape-changing and preventive maintenance.

\subsection{Observational Procedure} \label{observationalprocedure}

All observations are taken in a dark sky,  after the end of evening twilight and 
before the beginning of morning twilight, and with the moon below the horizon. A 
normal set of science observations consists of five 20-minute exposures through 
the H$\alpha$ filter interspersed among six 5-minute exposures through the dual-
band continuum filter. Use of multiple exposures allows removal of cosmic ray 
events and other transients, such as meteor, satellite, and airplane trails. 
Interspersing the H$\alpha$ and continuum exposures helps compensate for changes 
in sky foreground levels with time or zenith distance. 

Reflective ghosts of bright stars appear in the images. These are caused by 
light focused at a point on the CCD surface reflecting back through the lens to 
the filter, where it is again reflected and focused at a different point on the 
detector.  To assist in identifying and removing the reflective ghosts, the 
camera pointing is dithered in a raster pattern of 1/4$\arcdeg $ steps between 
each exposure. The ghosts move in the opposite direction to the stars themselves 
and so can be readily identified in the reduction procedure. (Transmissive 
ghosts are produced by small angle scattering off of imperfections in the 
filter. These move in the same direction as the stars during dithering, and so 
cannot be simply identified and removed during processing.)

Between observations of each science field, we take a few exposures pointed at 
the pole to obtain data on changes in extinction during the night. We also take 
occasional short exposures of bright standard objects. 

The program that schedules the observations gives priority to fields at large  
hour angle and high (more northerly) declinations, those for which the observing 
window is smallest.  Before observing more easterly fields it attempts to 
observe those fields which are about to become unavailable in the west. The 
program does not schedule an observation if the beginning or ending zenith 
distance would be greater than 65$\degr $. Because the camera is on a German 
mount, which must be flipped 180\arcdeg  at the meridian, observations are 
scheduled so they can be completed without crossing the meridian. Overall, 45\% 
of the fields were observed in the western part of the sky and 55\% in the east.

Biases, darks, and flats are taken during twilight at the beginning and end of 
each night. Each time we normally record 15 biases, five  darks each of 20 
minute duration (to match the maximum exposure times of the science images), and 
five flats for each of the two filters with the light box illuminated by two 
sets of incandescent lamps and for the H$\alpha $ filter with the light box 
illuminated by two sets of hydrogen emission tubes. The exposure times for the 
flats were adjusted to give images with a mean brightness several thousand times 
the readout noise, but below the level where the variance of the detector noise 
becomes non-linear.

\subsection{System Performance}\label{performance}

The system noise is a combination of photon noise from the sky foreground (with 
a small contribution from H$\alpha $ line emission), dark current noise, and 
readout noise. We can calculate the expected noise on a single exposure from the 
following formula:
\begin{equation}
N = {1 \over G} \sqrt{\lbrack {\beta ( I_{H\alpha } + S_{\lambda } \Delta 
\lambda)A \Omega T Q + D}\rbrack
\tau +{{\sigma }_{R}^2} }
\end{equation}
where N is the noise level (ADU), G is the gain (e ADU$^{-1}$), $\beta $ is the 
conversion factor of Rayleighs to photon intensity at 656.3 nm (${\rm{6.7 \times 
10^{-3}\ photons\ s^{-1}\ cm^{-2}\  arcminute^{-2}\ R^{-1} }}$), 
${\rm{I_{H\alpha }}}$ is the intensity in the H$\alpha $ line (R), 
${\rm{S_{\lambda }}}$ is the specific intensity of the sky foreground (R nm$^{-
1}$), $\Delta \lambda $ is the bandwidth of the filter (nm), A is the area of 
the lens (cm$^2$), $\Omega $ is the solid angle of the sky focused on one pixel 
(arcminute$^2$), T is the transmission of the optical system, Q is the quantum 
efficiency, D is the dark current (e s$^{-1}$), $\tau $ is the integration time 
(s), and ${\sigma }_{\rm{R}}$ is the readout noise (electrons). 

Using the parameters given in Section \ref{equipment},  a value of 10 R for the 
brightness of the geocoronal H$\alpha $ line \citep{noss93}, ${\rm{S_{\lambda 
}}}$ = 13 R/nm \citep{brkn68} for the specific intensity of the sky foreground, 
an assumed value of T = 0.6 (for the combined transmission of lens, filter, and 
CCD window), filter bandwidths of $\Delta \lambda $ = 3.2 nm and 12.2 nm for the 
H$\alpha $ and continuum filters respectively (Section \ref{filters}), and $\tau 
$ = 1200 s and 300 s for the integration times of the H$\alpha $ and continuum 
exposures (Section \ref{observationalprocedure}), we get expected noise values 
of 25 ADU  for a single H$\alpha $ exposure and  22 ADU for a single continuum 
exposure. Typical noise values actually achieved are 30 ADU and 24 ADU 
respectively, in reasonable agreeement with expectation. Note that for the above 
parameters the contribution to the noise from photons is larger than the 
detector readout noise, as desired, but only by a factor of two. The noise 
contributed by the dark current is completely negligible. 

The expected signal, after subtraction of the dark current, can be calculated 
from the formula:
\begin{equation}
S = {\beta \over G}   ( I_{H\alpha } + S_{\lambda } \Delta \lambda)A \Omega T Q 
\tau .
\end{equation}
It follows that the intensity scale factor $f$, the number of ADU corresponding 
to an intensity of 1 R, is given by 
\begin{equation}
 f = {S \over {( I_{H\alpha } + S_{\lambda } \Delta \lambda )}} = {\beta \over 
G} A \Omega T Q \tau .
\end{equation}
With the same parameters as used above for calculating the expected noise, we 
would predict an intensity scale factor of 11.2 ADU R$^{-1}$ for a 20-minute 
exposure. The actual value, determined from observation of planetary nebulae 
(Section \ref{intensityscale}) is 6.8 ADU R$^{-1}$, a difference of  40\%. One 
or more of the factors $A$, $\Omega $, $T$, or $Q$ may be overestimated, but if 
these estimates are reduced in order to generate better agreement between the 
expected and measured intensity scale factors, then the discrepancy discussed 
above between the expected and measured noise would increase.  Perfect agreement 
cannot be expected for such {\it ab initio} calculations, and we conclude that 
the system in the field performs reasonably close to expectations.

After co-adding the images (Section \ref{dataprocessing}), the noise is less. A 
typical value on the continuum-subtracted image (Section 
\ref{continuumsubtraction}) is 14 ADU (2 R), and on the images smoothed with a 5 
pixel median filter (Section \ref{smoothing}) it is 5 ADU (0.7 R). Large scale 
features of sub-Rayleigh intensity are visible on several images. We estimate 
that the faintest feature visible on our images has a brightness of 
approximately 0.5 rayleigh. 

Although the aim of this survey is the discovery and study of objects emitting 
at the H$\alpha $ line, there may be some applications in which the continuum 
point sources (stars) are of interest. The limiting R-magnitude on the co-added 
images is about 14.8 at b = $-90\degr $, deteriorating (because of confusion) to 
R = 12.5 at b = $20\degr $.

As mentioned in Section \ref{observationalprocedure}, we took short exposures of 
the south celestial pole region before and after each set of science images. 
Aperture photometry of several stars on these polar images showed that 
variations in extinction during one night or from night to night (for those 
nights for which data appears in the atlas) rarely exceeded 0.1 mag (10\%). 

Similarly (Section \ref{observationalprocedure}), we observed regions containing 
bright standard objects on most nights on which  data appearing in the atlas 
were obtained. We measured the brightness of carefully selected standard regions 
on these images. For those standard objects which were observed over a large 
range of elevations, we derived an average extinction coefficient of 0.082 mags 
airmass$^{-1}$ at 656.3 nm, compared to the standard extinction for CTIO at this 
wavelength of 0.096 mags airmass$^{-1}$ \citep{stone83}. The difference is 
within the seasonal and night-to-night variations of extinction at CTIO 
\citep{ham92,ham94}.  None of our data were obtained at zenith distances greater 
than $65\degr $, an air mass of 2.4 and extinction of 0.2 mag (20\%), and most 
images were observed much closer to the zenith. Again, individual intensities 
obtained on particular nights never differed by more than than 0.1 mag (10\%) 
from that expected for the zenith distance at the time of observation. 

Both the studies of the polar images and the standard objects indicate that 
variations in extinction do not affect the brightness of our images by more than 
10\%.  Therefore we have made no attempt to correct our data for atmospheric 
extinction.

\section{Data Processing} \label{dataprocessing}

\subsection{Calibrations} \label{calibrations}

Co-adding of the calibration exposures (biases, darks, and flats) is done on 
site automatically, with the medians of evening and morning sets recorded 
separately. These, and all science images, are recorded on tape and shipped to 
Swarthmore for later processing. No significant differences between morning and 
evening calibration images were detected, and these were simply averaged, if 
both existed.
 
The science images were corrected for bias and dark current in the usual way and 
then divided by the flat field image  appropriate to each filter, normalized to 
the  mean flat value of the central 100 pixel $\times $ 100 pixel area.

Early in the project we assumed that the lightbox was uniformly illuminated and 
emitted light at a constant brightness in all directions. We then attributed an 
observed decrease in brightness of the flats near their edges to vignetting in 
the camera system. However, using the flats as observed caused a rise in 
brightness of both the H$\alpha $ and continuum images near their edges, leading 
us to conclude that the lightbox emission was not truly isotropic, but decreased 
in intensity with angle to the normal. We do not know the origin of this effect, 
but suspect it may be caused by nonsymmetric scattering properties of the 
plastic diffuser screens. We corrected for this effect by determining the 
average of each flat image at constant radial distances from the center, and 
dividing the flat by this circularly symmetric function. Such a procedure 
preserves the features in the flat image due to pixel-to-pixel gain differences 
and to such things as dust specs on the camera window, but any symmetric 
features (such as true vignetting or gain variations which are a function of 
distance from chip center) are not corrected for and may propagate into the 
final images. However, examination of the continuum images at high galactic 
latitude ($|b| > 40\degr $), where there are no bright features, shows that on 
average they are flat to within $\pm3\%$ (1.3 R)  out to $8\fdg 0$ from the 
center (see Figure \ref{AvgDUPlot}), indicating that such vignetting-type 
effects, if they are present at all, are fairly small.

\subsection{Co-addition of Science Images and Removal of Ghosts and Transients} 
\label{coadd}

The sets of six continuum and five H$\alpha$ images are aligned and coadded 
separately, pixel by pixel, after removing transients produced by cosmic rays, 
meteors, airplanes, and satellites. We characterize a value as an outlier, to be 
excluded from the co-addition, if it lies more than 3.5 times the standard 
deviation from the mean. 

Given the nature of our data, determining the mean and standard deviation is not 
completely straightforward. First, since we have a small number of values 
contributing to each pixel (five or six), one of which (if influenced by a 
transient event) may differ by a large amount from the others, a straight 
average may not give a good estimate of the mean. Likewise, the rms average of 
the deviations from this mean may not give a good estimate of the standard 
deviation. Therefore we choose to take the median rather than the straight 
average as an estimate of the mean, and the median of the absolute  deviation 
from this median as an estimate the standard deviation.

A further complication is that each image is taken at a different zenith 
distance and therefore will contain a different airglow and geocoronal 
contribution to the foreground level.  The offsets due to these differences  are 
much larger than the noise. Therefore the median of pixels from the original set 
of images would give the value of the middle image, with no averaging of the 
noise. Likewise, the median of the absolute differences from this median would 
overestimate  the standard deviation, for it would include the offsets as well 
as the variations due to noise.

The first step, therefore, is to make corrections for the differences in 
foreground level. We assume that the continuum images consist of a galactic (and 
extragalactic) component $C_{g}$  plus a foreground component $C_{for}$ which 
varies with time:
\begin{equation}
C_{obs} = C_{for} (t)+ C_{g} .
\label{Cobs}
\end{equation}
Because the spatial transmission function of the H$\alpha $ filter (Figure 
\ref{HFilterTrans}) is still included in the line contribution to the H$\alpha $ 
images at this stage (it is removed later in the processing---see Section 
\ref{continuumsubtraction}), the corresponding relation for the H$\alpha $ 
images is a little different:
\begin{equation}
H_{obs}=1.2[C_{for} (t) + C_{g} ] +T_H [H_{for} (t) +H_{g} ]
\label{Hobs}
\end{equation}
where $H_{g}$ is the galactic  contribution to the H$\alpha $ line, $H_{for}$ is 
the time-variable foreground (geocoronal) contribution to the H$\alpha $ line, 
and $T_H$ is the spatial transmission function for the H$\alpha $ line. The 
factor 1.2 accounts for the different filter bandwidth and exposure times for 
the H$\alpha $ images compared to the continuum images, and is the average of 
the factors needed to match star intensities during the continuum subtraction 
process (Section \ref{continuumsubtraction}). Note: Equation \ref{Hobs} does not 
account for the contribution of telluric OH line emission (Section 
\ref{filters}) to the H$\alpha $ images, for which a different spatial 
transmission function should be used. To the extent that these lines are 
important, the separation of the line and continuum components of the H$\alpha $ 
image is only approximate.

We first determine the average brightness level, $C_{obs,n}$ or $H_{obs,m}$, of 
the central 800 $\times $ 800 pixel area of each image [using the "SKY" 
procedure of the IDL DAOPHOT package \citep[]{stet87,idl01}]. For the $n^{th}$ 
continuum image, we simply add a constant $C_{obs,1} - C_{obs,n}$  to bring it 
to the brightness level of the first image of the set. 
For the H$\alpha$ images, we use the function $C_{for}(t) + C_{g}$ determined 
from Equation \ref{Cobs} and from the $C_{obs,n}$ values to interpolate the 
first term of Equation \ref{Hobs}  to the time of each H$\alpha $ observation. 
Subtracting this term from the observed brightness level $H_{obs,m}$ gives  the 
line contribution to the brightness level, the second term of Equation 
\ref{Hobs}, and thus the function $H_{for}(t)$. It is then a simple matter, 
using Equation \ref{Hobs}, to correct each image to the brightness level at the 
time of the first image.
This procedure should compensate for linear changes in airglow and geocoronal 
level with observation time, at least within the time frame of a particular set 
of images. (In Section \ref{georemoval} we discuss a procedure for matching 
levels of {\it all} images.) Since all the images of the set now have the same 
average value, medians of the image values at each pixel and medians of the 
absolute deviation at each pixel should give a good estimate of the mean and 
standard deviation.

Next, reflective ghosts are identified and marked on each image. This procedure 
could not be completely automated because of a random variation by several 
pixels of the position of the axis of reflection from image to image. This is 
probably due to an imperfect constraint of the plane of the filter wheel or a 
slight looseness of the filters in their mounts. 

An empirical study of a few images showed that the ghosts are about 0.05\% of 
the brightness of the stars, so that only the very brightest stars, those with 
central brightness exceeding the full-well capacity of the chip (\rm 80,000 e 
$\approx $ 60,000 ADU), produce ghosts on an image greater than the noise. In 
the ghost-removal procedure, an approximate axis position is assumed, the 
positions of the ghosts of the brightest stars calculated and marked on the 
image display, and the axis position adjusted interactively until the marked 
positions coincide with the real ghost positions. A circle of radius 12 pixels 
($9\farcm 5$) around each ghost is marked and not used when the images are co-
added.

Next the images are approximately aligned by shifting and rotating to remove the 
positional dither. They are then aligned more exactly by finding the positions 
of all the intermediate brightness stars, those with central intensity between 
10 times the sky foreground and a level of 14000 ADU, and then finding the 
shifts and rotation which minimize the differences of position between the stars 
on each image compared to the first one of the set.

The aligned images are then co-added after removal of cosmic ray events and 
other transients. As described above, this is done by first finding at each 
pixel position the median of the set of images, and the median of the absolute 
difference from this median. For small numbers of points, the mean can be 
significantly different from the median, so now the true mean at each pixel 
position is  calculated, excluding those points whose difference from the median 
is more than 3.5 times the median absolute difference. Regions containing ghosts 
are also excluded when calculating the mean. 

\subsection{Continuum Subtraction} \label{continuumsubtraction}

At this point we have co-added means of the images taken through the H$\alpha $ 
and continuum filters, with transients and ghosts removed. Next the continuum 
image is aligned to the H$\alpha $ image, using the same procedure as described 
above. Aperture photometry on the same intermediate brightness stars on each 
image establishes an intensity scale factor between the two mean images. The 
continuum image, multiplied by this scale factor, is then subtracted from the 
H$\alpha $ image. The continuous airglow foreground and most stars are removed 
by this procedure, but the geocoronal H$\alpha $ emission and  residuals of the  
bright stars remain. These bright star residuals are probably due to slight 
differences of the point spread functions on the two images,  small non-
linearities of the CCD response, or saturation of the detector (for the very 
brightest stars). A sample of the continuum,  H$\alpha $, and star-subtracted 
images for one field is shown in Figure \ref{OrionQuad}.

The image should in principle now contain only line emission, primarily H$\alpha 
$, both interstellar and geocoronal, and stellar H$\alpha $ emission and 
absorption, as modified by the transmission function of the H$\alpha $ filter. 
(There remains also some residual contamination by telluric OH lines---Section 
\ref{filters}.) The basic transmission function is determined by dividing the 
H$\alpha $ flat (produced by observing the lightbox illuminated by the hydrogen 
emission lamps) by the continuum flat (produced by observing the lightbox 
illuminated by the incandescent lamps), both with the H$\alpha $ filter in 
place, and both normalized to the mean of the central 100 pixel $\times $ 100 
pixel area. In principle the H$\alpha $ component of the original images should 
be corrected for this transmission factor before they are shifted and co-added, 
but it is difficult to separate accurately the line and continuum fractions of 
the image before the co-addition. Therefore we constructed an effective H$\alpha 
$ transmission function by de-dithering and co-adding the original function in 
precisely the same way as was done for the original images obtained with the 
H$\alpha $ filter. The star-subtracted image was then divided by this function 
to obtain a corrected image, with the transmission characteristics of the 
H$\alpha $ filter removed.

\subsection{Corrections} \label{corrections}

For many fields, this procedure produces a good image, planar except for real 
interstellar H$\alpha $ features. For some fields, however, obvious instrumental 
effects remain. One of these manifests itself as an increase of brightness with 
radius, in shape like the inverse of the H$\alpha $ filter transmission function 
(Figure \ref{HFilterTrans}). This probably arises from imperfect subtraction of 
the continuum, perhaps due to the foreground brightness changing non-linearly 
with time. If some continuum radiation remains in the image, then division by 
the H$\alpha $ filter transmission function will cause such an upturn near the 
edges. To correct for this phenomenon, we iteratively subtracted from the images 
showing this effect a constant divided by the H$\alpha $ filter transmission 
function until we found the value which minimized the upturn near the edges.

Even with this effect removed, some images showed a residual pattern at low 
amplitude (a few rayleigh) of a circularly symmetric brightness, rising and then 
falling with radius. Such an effect could arise from one or more emission lines 
in the night sky spectrum shifting across the filter band with increasing angle 
of incidence. But the position and shape of the maximum does not seem to 
correspond with that expected for any known line. It could also be some artifact 
of the flat-fielding procedure. The cause of this effect is  not understood.

We determined this  function individually where we could, for those images which 
contained no bright interstellar H$\alpha $ features, by finding the average of 
each image at constant radial distances from the center, after normalizing the 
image to have zero mean, and subtracting this circularly symmetric function from 
the image. For those fields, mainly at low galactic latitudes, where the genuine  
H$\alpha $ features were too bright for this procedure to work, we merely 
subtracted the average of all the functions determined at high latitudes.

An example showing graphs of the  average brightness as a function of radius for 
an original continuum-subtracted image and for the same image after these two 
stages of correction is shown in Figure \ref{curlsequin}. 

\subsection{Smoothing} \label{smoothing}
These procedures produced images which are fairly planar, but they do contain 
geocoronal H$\alpha $ emission (and possibly some residual OH emission) as well 
as H$\alpha $ emission from the interstellar medium. The geocoronal emission is 
variable in time and with position on the sky. We have assumed that it changes 
fairly slowly in both dimensions, and have modelled it as a plane with a certain 
level (which we call a piston), slope in declination 
(tip), and slope in right ascension (tilt). We determined these parameters 
(piston, tip, and tilt) for each image so that neighboring fields fit together 
into a smooth mosaic, as described in Section \ref{georemoval}.

Finally, as mentioned above, the star-subtraction procedure does not work 
perfectly, so that the images contain some high spatial frequency residuals of 
imperfectly subtracted bright stars. We removed these as best we could, but with 
some loss of resolution, by applying a spatial median filter of width 5 pixels 
(4\farcm0) to the images. The smoothed image for the Orion region is shown in 
Figure \ref{OrionQuad}d.

\subsection{Astrometry} \label{astrometry}

By automated pattern-matching of stars in our images to
those in the SAO catalog, we determine astrometric solutions
for each mean \halpha~and mean continuum image. The solution,
a gnomonic projection, is stored in the image header according
to standard practice \citep[]{cala00}.
Here we detail the method of solution and describe its accuracy. The method
is similar to others that we might have used (e.g., \citet[]{mink97})
if they had existed when we first needed them.

After each observation, 
the telescope control software annotates the image header
with the celestial coordinates of the intended center of the image.
From the image, our algorithm generates a list of stars' centroids
and instrumental magnitudes.
From the SAO catalog, our algorithm culls a list
of J2000 positions and apparent visual magnitudes of
stars within 9\arcdeg~of the intended center. The stars' celestial
positions are converted to angles $(\xi,\eta)$ of a gnomonic projection
centered on the image's intended center (e.g., \citet[]{mars82}).
The pattern-matching algorithm compares the list of instrumental
centroids and magnitudes to the list of gnomonic angles $(\xi,\eta)$
and SAO magnitudes. It uses  Valdes' method of similar
triangles \citep[]{val95} as implemented by \citet[]{rich96} to determine
a tentative match between the 50 brightest stars in one list and the
50 brightest stars in the other list.
It then iteratively refines the solution by extending the star
lists, typically to 200 stars in common.
The custom keyword N\_ASTROM in the image header logs the number of stars
that contributed to the astrometric solution.

The instrumental positions match the
catalog positions to $0.18$ pixel (8\arcsec) r.m.s. if first-order
plate equations are applied. The residuals of the first-order fit exhibit
a third-order polynomial with maximum amplitude of ${\pm}0.4$
pixels (${\pm}0\farcm32$). Fitting with third-order polynomials
reduces the r.m.s. of the residuals by a factor of two,
but for simplicity we chose to retain only
the first-order fit in the image header.

Astrometric solutions were computed separately for the co-added H$\alpha $ and 
continuum images. 
Because the continuum image was shifted to match  the H$\alpha $ image 
prior to subtraction, the continuum-subtracted images and the smoothed images 
have the same astrometric
solution as the H$\alpha $ image.

\section{Intensity Scale} \label{intensityscale}

To calibrate the intensity scale of our images, we compared 
aperture photometry of many compact (diameter $< 5$\arcsec) planetary nebulae 
(PNe) on
our images with the spectrophotometric fluxes reported by \citet[]{dop97}
(see Figure \ref{dopitaplot}).
Dopita \&  Hua report fluxes for \halpha~(656.3 nm)
and the forbidden line of [N II] at 658.34 nm. The latter spectral line and its
companion at 654.81 nm contribute to the flux in our ``\halpha'' images.
The intensity of [N II] 654.81 nm
is 1/3 that of [N II] 658.34 nm \citep{ostr89}.
The transmission of the ``\halpha'' filter,
at our 25 \degr C operating temperature and at normal incidence\footnote{The
filter's transmission at 654.81 nm increases with increasing angle of
incidence, while at 658.34 nm the transmission decreases with increasing
angle of incidence(Figure \ref{FilterTrans}). The net effect is that the
intensity-weighted average transmission of the two [N II] lines is a weakly
varying function of angle of incidence that we have not attempted to allow
for in determining the overall intensity calibration.},
is 39\%, 78\%, and 26\% at 654.81 nm, 656.3 nm, and 658.34 nm respectively.
By accounting for the expected relative contributions to the flux measured
through our filter for each PN according to the fluxes given by Dopita
\&  Hua, we determined the single scalar that best matched our instrumental
fluxes (in analog to digital units, or ADU) to theirs (in \ergscmsec).
We chose to use in the comparison group
only those PNe  that had an \halpha~flux larger than
$10^{-11.25}$ \ergscmsec , because many of the fainter ones had
large uncertainties in our aperture photometry. If the aperture photometry
routine [APER in the IDL DAOPHOT package \citep[]{idl01}] reported an error of 
less than 6\% for any
particular measurement, we replaced that error estimate with 6\%; three
such replacements occurred in each of the two groups (open or filled
circles) plotted in Figure \ref{dopitaplot}.
For those 18 PNe selected for the intensity calibration,
the r.m.s. differences between our \halpha~fluxes, determined from
continuum-subtracted images and corrected for [N II] emission,
and the \halpha~fluxes of Dopita \&  Hua, is 8\%. Because
Dopita \&  Hua's spectrophotometry is accurate to better than $1$\%,
we consider the 8\% r.m.s. scatter between our measurements and
theirs as indicative
of the limit of precision of fluxes of unresolved sources measured from
our continuum-subtracted images. We attribute this 8\% limit ($1\sigma$)
on precision  to uncorrected  extinction (Section \ref{performance}) and 
the undersampling of the optical system's point spread function by our pixels.
The transmission of our ``continuum'' filter
at the wavelength of the [SII] line at 673.1 nm is 16\% at normal
incidence, and as large as 70\% at the corners of our images (\ref{filters}), 
but
because our continuum images are exposed for 25\% of the time of the
\halpha~images, and because
in all of these PNe the [S II] $\lambda$673.1 line is
less than 10\% the brightness of \halpha, we have not corrected our
fluxes for [SII] as we did for [N II]. As a worst-case example,
a single measurement
of a particular PN that happened to be near the corner of an image,
the leakage of [S II] $\lambda$673.1 would amount to
less than 2\% of the \halpha~flux. In an image in which that same PN 
appears near the center, the effect would be less than 0.4\%.
Using the PNe calibration just described, and converting from flux
to surface brightness units, we find that in a single pixel of
our \halpha~images, 1 R = $6.8 \pm 0.3$ ADU. This factor has
been applied to all the continuum and smoothed images available 
for public use (Section \ref{access});   
their intensities are expressed in units of decirayleighs (1 dR = 0.1 R).

As an independent check of the calibration using the 18 PNe described
previously, we compared our measurement of the total \halpha~flux of the
very bright Rosette nebula to the \halpha~flux measurement of \citet{cel83}.
Integrating the flux within a 1\arcdeg~radius circle, as Celnik did,
and setting our
image's zero point to agree with the dashed contours of Celnik's Figure 2,
and correcting our measurements for [N II] leakage with the spectral line
ratio measured for the Rosette,
I(6583)/I(\halpha) = 0.37 (\citet{john53}; position 58) we find 
1 R = $7.4 \pm 0.5$ ADU, where the uncertainty is set by the
7\% uncertainty reported by Celnik for the integrated \halpha~flux.
Because the PNe calibration is based upon many objects measured on many images
rather than a single object (the Rosette) measured on a single image, 
the Rosette calibration is secondary to 
the PNe calibration that we actually applied, 1 R = $6.8 \pm 0.3$ ADU.
The ratio of the two (7.4/6.8 = 1.09) indicates that our intensity
calibration is consistent with others to approximately $\pm 9$\%.
(See Section \ref{others} for comparisons with other surveys.)

\section {Removal of Geocoronal Emission} \label{georemoval}

Each individual \halpha~image\footnote{Throughout this Section, ``image'' means 
an image that has been
completely calibrated according to the description in the previous sections,
unless specified otherwise.}
has an unknown offset due primarily
to geocoronal \halpha~emission, which is spatially and temporally variable.
To account for variations in geocoronal emission,
we have adopted a model in which the extra foreground emission is
a 2-dimensional (2-D) plane, the three parameters of which we call ``piston''
for the mean value of the intensity of the foreground (\barF) and
 ``tip'' and ``tilt'' (collectively \gradF) for the foreground's slopes.
Thus, the galactic emission, $I^g$, that we seek can be determined 
from the emission that we observe, $I^o$, as follows:
\begin{equation}
I^g = I^o - (\bar F + \nabla F \cdot \vec X),
\end{equation}
where $\vec X$ is a linear transform of the x- and y-coordinates of each pixel:
$\vec X=[(x-<x>)/<x>,(y-<y>)/<y>]$.

The challenge is to find the specific \barF~and \gradF~ for each
individual image in a physically justified manner. 

Our approach to this challenge extends
the method of \citet{r&g95} in a few ways:
1) theirs solves for \barF, while ours solves for \barF~and \gradF;
2) theirs matches pixels in one image with pixels in an
adjacent image using a simple shift in $\vec X$, while ours uses the astrometry
stored in the FITS headers;
3) theirs uses a median of differences of matched pixels in each
overlapping region, while ours fits a 2-D plane to those differences;
4) the overall zero-point(s) for the pistons (and tips and tilts in our
case) are determined by different methods;
5) their algorithm for solving the system of equations is
a simplex method; ours is a singular value decomposition.
The large angular size of our images required us to implement item
1 because the foreground sky is not constant across an image,
item 2 because of image distortion and rotation,
especially near the celestial pole, and item 3 because according to
our model the difference between two adjacent images is a 2-D plane
whereas in their model the difference is a single scalar.

Our method of determining the specific \barF~and \gradF~ for each
individual image is best described by an example. Figure \ref{4images}
shows four overlapping images. There are five pairs of overlapping
images:  images 1 and 2, 1 and 4, 2 and 3, 2 and 4, plus 3 and 4
overlap. We begin with the overlap between image 1 and 2. We subsample
the ``good'' part of image 2 on a 20 pixel by 20 pixel grid and compute
the position on the sky ($\alpha,\delta$) of each pixel from image 2's
astrometric solution. For those ($\alpha,\delta$) we compute the (x,y)
in image 1's astrometric frame in order to pair the pixels in image 2
with pixels in image 1 at the same ($\alpha,\delta$).  Each such pair
of pixels provides an equation such as
\begin{equation}
I^o_1 - I^o_2 = (I^g_1 + \bar F_1 + \nabla F_1 \cdot \vec X_1) - (I^g_2 + \bar 
F_2 + \nabla F_2 \cdot \vec X_2),
\end{equation}
where the subscripts refer to the image (1 or 2) and the superscripts
($o$ or $g$) refer to observed or galactic intensity.
Because the galactic emission is  constant with time, i.e.,
$I^g_1 = I^g_2$, it cancels out in the difference and we have
\begin{equation}
I^o_1 - I^o_2 = (\bar F_1 + \nabla F_1 \cdot \vec X_1) - (\bar F_2 + \nabla F_2 
\cdot \vec X_2).
\label{i1mi2}
\end{equation}
In this equation we have observed $I^o_1$ at location $\vec X_1$ of image 1
and have observed $I^o_2$ at $\vec X_2$ of image 2. The unknown parameters
for images 1 and 2 are
$\bar F_1, \nabla F_1, \bar F_2,$ and $\nabla F_2$.

Likewise, each pixel in image 3 that overlaps image 2 provides an equation
\begin{equation}
I^o_2 - I^o_3 = (\bar F_2 + \nabla F_2 \cdot \vec X_2) - (\bar F_3 + \nabla F_3 
\cdot \vec X_3).
\end{equation}
and similarly for the three remaining overlapping image pairs.

We can define a matrix equation $\vec b = \matrixA \cdot \vec v$ where
vector $\vec b$ contains the intensity differences between all pixel pairs of
image i and image j, i.e., $\vec b = \lbrace I^o_i - I^o_j \rbrace$;
vector $\vec v$ contains the model parameters, 
i.e., $\vec v = \lbrace \bar F_i, \nabla F_i \rbrace$; and
\matrixA~describes the coupling of $\vec b$ and $\vec v$
by equations like those above. 
In the illustrated example, $\vec v$ would be 12 elements long (4 images and
3 parameters per image).
Each pair of pixels from adjacent images adds an element to $\vec b$
and a row to \matrixA.

We then solve $\vec b = \matrixA \cdot \vec v$ for $\vec v$ by a standard
method of linear algebra, singular value decomposition.  However, even 
if we have more equations
than unknowns, the equations are degenerate such that the overall
piston, tip, and tilt are arbitrary. In many circumstances it may
suffice to append three rows to \matrixA~to force the mean values
$<\bar F_i>$ and $<\nabla F_i>$ to zero.  Another option is to
refer the data to intensity values known by other means.  In our case
we get additional equations by comparing our data with WHAM data.

If all calibrations are done correctly,
 if we average our data over a 1\arcdeg-diameter circle
of sky and integrate WHAM's spectra over velocity while eliminating 
the geocoronal emission in the spectra, and if the contamination of our data
by telluric OH emission is negligible, then
our measurement of the galactic \halpha~intensity
and WHAM's should be equal, i.e., $<I^g> = I_{WHAM}$. 
Three WHAM data points are shown in Figure \ref{4images} as circles. 
In our hypothetical example, we append to \matrixA~equations like the
following:
\begin{equation}
<I^o_3(c)> - I_{WHAM}(c) = (\bar F_3 + \nabla F_3 \cdot \vec X_3).
\label{eq-Iwham}
\end{equation}

In this equation, which applies to point c in Figure \ref{4images},
$<I^o_3(c)>$ is the observed intensity of image 3 centered on point c and
averaged over a 1\arcdeg~diameter
circle that corresponds to the WHAM beam size; $I_{WHAM}(c)$ is the
intensity at point c as measured by WHAM; and $\vec X_3$ is the
location of point c in image 3's coordinate system.
In principle, only three such fiducial points (e.g., points a, b, and c)
are needed to constrain the overall piston, tip, and tilt. 
We append to \matrixA~equations similar to the one above for image
2 and point a, image 3 and point b, and image 4 and point c, but
not image 2 and point b because the circle around point b does not lie
wholely within image 2's ``good'' region, making $<I^o_2(b)>$ inaccurate.

For small numbers of images, it is practical to simply use a very
sparsely subsampled grid of pixels, and let every pair of pixels
lengthen $\vec b$ by one element and \matrixA~by one row.
Because the time to solve the equation $\vec b = \matrixA \cdot \vec v$
for $\vec v$ grows in proportion to the product of the number of pixel
pairs times the square of the number of free parameters of the model
(in our case, three times the number of images), a more efficient
approach is required for large numbers of images.  We chose to
subsample our images, and then in each region of overlap, we inscribed
a rectangle and within the rectangle we fit a 2-D plane to
the differences in intensities of paired pixels. We then evaluated the
plane at the corners of the rectangle and used those values as
representative of all the paired-pixel differences and locations in the
region of overlap.  Choosing the rectangle's four corners of course
over-determines the plane, but doing so is a simple and approximate
method of representatively weighting the data as it would be if all
pixels within the irregularly-shaped region of overlap were given equal
weight.  An iterative method is used to fit the plane robustly, i.e.,
such that outliers do not affect the solution.

Simulations in which a large image is carved into a set of overlapping
images and then random pistons, tips, and tilts are added to those
images, and subsequently solved for, show that our algorithm can
recover the pistons, tips, and tilts very accurately and reliably under
optimum conditions.  In practice, with our actual \halpha~images, our
experience with the algorithm has been much less satisfying.

With our
\halpha~images, if we solve for \barF~and \gradF~simultaneously
the solution tends to introduce large scale undulations
in the mosaic of the (galactic) \halpha~sky. Wherever the solution is
not tied to  fiducial points, such as in the southern
sky ($\delta < -30\degr $) where no WHAM data exists, the solution tends
to be unreliable.   Slightly different input parameters give rather
different results, even physically impossible results, such as negative 
brightness in
some regions.  The solution tends to err in such a way that in
extrapolating beyond the set of images containing WHAM fiducial points to those
without any, erroneous gradients trade off with erroneous pistons, so
that the solution is ``smooth'' from one image to another, but  has
the wrong overall ``figure'' or shape.
We tested the reliability and accuracy by using a subset of the WHAM
data to determine a solution and a different subset to compare with our
data after the solution has been applied. For example, to test our
algorithm's ability to extrapolate from the north to the south, i.e.,
from where there is WHAM data to where there is not, we included as
input (cf. Equation \ref{eq-Iwham}) WHAM data with declination $\delta
\ge -5$\arcdeg, determined a solution, applied it to our images, and then
compared our resulting \halpha~intensities with WHAM data in the region
$-25$\arcdeg~$~\le~\delta~\le~-10$\arcdeg. This empirical test
convinced us that in order to prevent the solution from diverging in
the southern sky where no WHAM data exists, i.e., $\delta < -30$\arcdeg,
we had to provide additional, artificial fiducial points in that region
that would be treated the same as actual WHAM observations.

In order to create these artificial fiducial points, which are shown in Figure 
\ref{figwhamaitoff}, we  fit a function of the form $C_1 + C_2 / sin |b|$ to 
WHAM data with $-25\arcdeg \le \delta \le +10\arcdeg $ and
$|b| > 30$\arcdeg, but excluding the Orion/Eridanus region,
$150\arcdeg  < l < 230\arcdeg $. The parameters $C_1$ and $C_2$ were
fit by eye in order to ignore the somewhat-brighter regions remaining 
within those limits of $\delta$, $b$, and $l$. The function so selected,
\begin{equation}
I^g(model) = -0.75 + 1.0/sin|b|~{\rm [rayleighs]},
\label{cscblaw}
\end{equation}
is plotted in Figure \ref{figcscblaw}. 
That the \halpha~sky is fainter at high latitudes than the extrapolation
of a $csc|b|$ law determined at lower latitudes is believed to be due
to the Local Bubble, plus a small, increasing contribution for dust-scattered 
H$\alpha $ 
at the lower latitudes \citep[]{wood99}.
If we assign all of this effect to the Local Bubble, then the isotropic 
component in Equation \ref{cscblaw},
-0.75 R, corresponds to the emission missing from that part of the WIM
that the Local Bubble has ``carved out.''
For a mid-plane r.m.s. electron density 
$\sqrt{<n^2_e>} = 0.084$~cm$^{-3}$ \citep[]{rey91},
the ``missing'' \halpha~intensity of 0.75 R requires a 240-pc path length
of fully ionized gas at 8000 K.
That length is approximately equal to the 200-pc radius of the Local Bubble
as modeled by \citet[]{wood99}, so our paramter $C_1 = -0.75$ rayleigh is
consistent with prior models for the local WIM.

In the end, we chose to include data from all 542 images to determine
simultaneously the \barF~and \gradF~for each of the images. 
With 3166 pairs of overlapping images, and each region of overlap
represented by 4 points as previously described, there were
12664 equations like Equation \ref{i1mi2}.
We used 802 WHAM reference points spaced 
4\arcdeg~in right ascension between $0^h \le \alpha \le 24^h$, and spaced
5\arcdeg~in declination between $-25\arcdeg \le \delta \le +10\arcdeg$.
We did not use any
WHAM-measured surface brightnesses less than zero (because those
are probably erroneous) or larger than
30 R because we wanted to avoid bright
regions for which multiplicative errors might affect our estimate of the
additive terms. We also used 527 artificial points created as
described previously (cf. Equation \ref{cscblaw}).
The 802 WHAM fiducial points and 527 artificial fiducial points
have 4803 separate measurements in our set of 542 images,
resulting in 4803 equations like Equation \ref{eq-Iwham}. 
In order to give approximately equal weight to the two types of
equations, i.e., those that compared overlapping images
(Equation \ref{i1mi2}) and those that compared our data to WHAM data
(Equation \ref{eq-Iwham}), we replicated the latter 4803 equations
to produce 9606 such equations, or nearly as many as the 12664 equations
of the other type. We then solved the equations for the 542 images'
\barF~and \gradF~by singular value
decomposition. We solved the equations twice; prior to the
second iteration, we discarded 52 (0.4\%) of the 12664 equations that
were found to deviate by more than 10 times the
average residual difference after applying the initial
solution's \barF~and \gradF. As expected, because only a very small
fraction of the equations were excised, the second set differed only
very slightly from the first set.
The pistons, tips, and tilts determined in the second set were applied to
the continuum-subtracted (and corrected) images and the smoothed images 
(Sections \ref{continuumsubtraction}--\ref{smoothing}), and the latter were used 
to make 
the mosaic map mentioned in Section \ref{description}.

\section{The Survey}\label{survey}

\subsection{Description and Sample Images}\label{description}

The survey consists of 268 fields covering the sky from $-90\degr $
to $+15\degr $ declination,
with the same centers as those in the IRAS Sky Survey Atlas 
(ISSA)\citep[]{whe94},
 plus,  instead of ISSA region 001, three fields centered at $-87\fdg 5$ 
declination and 0, 8, and 16 hours right ascension to cover the polar region. In 
order to reduce
the noise in mosaic maps and to allow confirmation of objects discovered at the 
faintest brightness levels,
we repeated all fields with offsets of  $5\degr $ in both coordinates. (The 
three polar fields were repeated at the same declination, $-87\fdg 5$, but 
displaced 4 hours in right ascension.) 

The field centers are listed in Table \ref{CentersTable}.

Images of the individual fields are available from our web site (Section 
\ref{access}).
Some examples of interesting fields are shown in Figure \ref{favorites},
and a mosaic of the entire region of the sky included in the atlas is shown in 
Figure \ref{mosaicfig}.

All observations used in the survey were obtained in the period November 1997 to 
October 2000, a time of increasing solar activity.

\subsection{Comparison with Other Surveys} \label{others}

\subsubsection{VTSS, the Virginia Tech Spectral-Line Survey} \label{vtss}

The Virginia Tech Spectral-Line Survey (VTSS)
is an imaging survey similar to the one described here 
but for the northern
hemisphere \citep{dst98}. 
When the VTSS observations of the northern hemisphere are complete,
combining them with our southern hemisphere data will provide
an all-sky \halpha~mosaic with approximately 3\arcmin~resolution.
Comparison of VTSS data with our own shows they are very similar,
but with some differences, as the remainder of this section describes.

The VTSS includes much of the Milky Way ($|b| \la 30$\arcdeg)
clearly visible ($\delta \ga -15$\arcdeg)
from the Martin Observatory in southwestern Virginia.
The VTSS's cryogenically-cooled CCD camera has a 58-mm lens, 27~\micron~pixels,
a scale of 1\farcm6 pixel$^{-1}$, and covers a
13\fdg6 $\times $13\fdg6 field of view. The 
1.75 nm bandwidth (FWHM) of the VTSS's \halpha~filter limits
the usable field of view (cf. Equation \ref{cwlshift})
to  a 10\arcdeg~diameter circle centered on the optical axis.
As soon as they are processed to final form,
individual images from the VTSS are posted on the VTSS
web site (\url{http://www.phys.vt.edu/\~{}halpha/}).

In order to make the comparison, we selected four 
continuum-subtracted \halpha~images from the VTSS:
Aql04, Aql12, Cnc02, and Ori11. [The first three letters of each image's
name refer to the constellation (Aquila, Cancer, and Orion) and the number
is simply a counter.] Each of the first three VTSS images is an average of
six 360-second exposures, and the last image (Ori11) is an average of
twelve 600-second exposures.

We mapped, pixel-by-pixel, all of our final smoothed images that overlapped
each VTSS image onto that VTSS image. 
Because our pixel size is 0\farcm8 and the VTSS images' pixel size is 
1\farcm6, each of our images has four pixels within the area of one VTSS pixel.
Between seven and ten of our images
overlap each VTSS image, but most of them do so only partially, such that
typically two to four of our images cover a particular point on the sky.
Thus, in mapping our images upon a VTSS image, we average between 8 and
16 pixels' data for each VTSS pixel. The resulting mosaic of our images,
custom-made to match the pixel scale and orientation of the VTSS image,
is then nearly ready to be compared with the VTSS image.

Because a constant has been subtracted from each VTSS image to force its
median to be zero, we need to correct the zero-point of intensity scale
of each VTSS image.
Rather than calculate a simple offset, we fit an arbitrary 2-D plane of
emission to the difference between the VTSS image and our custom-made
mosaic. By subtracting that 2-D plane from the VTSS image, we
effectively corrected the VTSS image for whatever geocoronal 
or telluric OH foreground emission
might be included in it. (Our images have been corrected already
for this emission as described in Section \ref{georemoval}.)

At the top of Figure \ref{fig-vtss},
we show the VTSS image Aql04 and our custom-made
mosaic, for side-by-side visual comparison.\footnote{Actually we have
also corrected the grey scale of the VTSS image Aql04 by dividing it by 0.80,
a factor that will be explained shortly.}
Notably, the structure of the diffuse emission
is nearly identical, as it should be, because both of us are measuring the
\halpha~emission.
The VTSS image has more noticeable residuals around stars, but aside from 
that the two images are nearly identical. The width of the
point spread function, measured from the same few unresolved emission knots,
is $\sim 2$ pixels (3\farcm2) FWHM in either image.
In the VTSS images, this FWHM
represents an intrinsic resolution limit, while in our case it represents
a kind of effective width of the native resolution of our images
and the $4\arcmin \times 4\arcmin$ kernel of the
median filtering of our images, combined with the binning and
averaging of multiple images that make up each custom-made mosaic.

Also in Figure \ref{fig-vtss}, in the lower left panel, we show a pixel-by-pixel
comparison of the VTSS image Aql04 and our mosaic image, after each has been
convolved with a $5\times 5$ pixel ($8\arcmin \times 8\arcmin$) kernel. 
The convolution serves only to reduce noise and to reduce effects of
slightly dissimilar point spread functions. Rather than plot {\it every}
pixel, which would make the density of points in the plot too high,
we randomly selected 1\% of the pixels across the image for the plot.
From this plot, and especially the one next to it, from the Ori11 region,
we determine that the intensity calibration
of our data and the VTSS data are not the same. For an object
within the Ori11 region,
the surface brightness measured from a VTSS image will be 80\% that 
measured from our images. For the Aql04 region the measured value is 85\%,
but with greater uncertainty due to the smaller dynamic range of
surface brightness in Aql04 compared to Ori11. For the other two images,
Aql12 and Cnc02, the \halpha~emission is too faint for the intensity ratio to
be determined accurately.
For display purposes, we have corrected the
VTSS image Aql04 in Figure \ref{fig-vtss} by dividing it by 0.80, but we have
not attempted to determine the cause of this effect or whether it affects
other VTSS images or perhaps only these two images that we have examined.

In order to compare the sensitivity of our survey to that of the VTSS,
we further processed the images of the two regions Aql12 and Cnc02,
which have some faint emission in them, by median filtering each with a
$0\fdg4 \times 0\fdg4$ kernel and subtracting the result from each
image in order to create residual images. The purpose of creating
residual images is to prevent genuine emission from affecting
our measurement of the noise in the image. The r.m.s. surface brightness
in those residual images were 0.23 R and 0.15 R for VTSS images Aql12 and Cnc02,
respectively\footnote{If the factor of 0.80 determined from the Ori11 field
should apply to the Aql12 and Cnc02, then the values 0.23 R and 0.15 R
should be adjusted to 0.29 R and 0.19 R, respectively.},
and 0.22 R and 0.22 R respectively for their associated
custom-made mosaics from our data. In determining the r.m.s., we used only
the cores of the distributions, not the wings that are created by the
residuals around stars. Thus, except for the possible difference in
overall intensity calibration, when convolved to a similar resolution,
our data and the VTSS data give
similar results in those regions of sky in which they overlap (i.e.,
for $-15\arcdeg \la \delta \la +15\arcdeg$, and $|b| \la 30\arcdeg$).

\subsubsection{WHAM, the Wisconsin H-Alpha Mapper survey} \label{wham}

As noted in Section \ref{intro},
the WHAM survey is a velocity-resolved survey
of H$\alpha$ emission over 3/4 of the entire sky ($\delta \ga -30$\arcdeg).
The WHAM survey consists of more than 37,000 individual
spectra, each with velocity resolution of 12 km~s$^{-1}$ over a 200 km~s$^{-1}$
(0.44 nm) spectral window.
Each velocity-integrated spectrum has a 3$\sigma$ sensitivity of 0.15 R
over its 1\arcdeg~diameter beam. 
Geocoronal \halpha~emission is separated from Galactic \halpha~emission
spectrally in the WHAM data.
Because of the WHAM survey's sensitivity and accuracy, it is an ideal
data set to compare our own to, after appropriately integrating the
WHAM spectra and smoothing our data to match WHAM's 1\arcdeg~angular
resolution.

Figure \ref{hassmwham} compares WHAM data with our own.
In Figure \ref{hassmwham}, the WHAM values are velocity-integrated and have 
had geocoronal emission removed \citep{rey01}, while
this work's values are the average within the 1\arcdeg~diameter
beam of WHAM of our final smoothed images after removal of geocoronal emission 
as described in Section \ref{georemoval}.
The locations on the sky of the 834 points plotted in Figure \ref{hassmwham}
are shifted 2\arcdeg~in right ascension with respect to the  locations of
the points used to fit and remove the geocoronal emission, 
in order to assure that the set of points used for the comparison
contains none of the points used for the fitting.

The r.m.s. of the residuals
of points with WHAM-measured surface brightness less than 2 R is 0.4 R.
Because the WHAM sensitivity is 0.05 R ($1\sigma$), the 0.4 R r.m.s.
difference between WHAM's values and our own is almost entirely
due to our data's noise and is an empirical limit of the sensitivity
of our data after convolution with a 1\arcdeg-diameter beam.
For faint sources, the dominant errors are additive, but for bright
sources, multiplicative errors could be more important than additive
ones.
In fitting and removing foreground emissions (Section \ref{georemoval}),
we specifically excluded regions brighter than 30 R
so that
any multiplicative errors that might exist would not affect the fitting
too badly. Consequently, for a fair comparison of our intensity scale to
WHAM's, we should include only regions much brighter than 30 R in order to 
reduce the effect of the additive offsets determined by the fitting procedure.
Generally, bright regions are better than faint regions for comparisons of two
independent scalings of intensity.
For the 15 points in Figure \ref{hassmwham} with WHAM-measured
surface brightness greater than 60 R, the ratio of our values to
WHAM's has a mean of 1.00 and a $1\sigma$ dispersion of 0.13, 
i.e., our intensity scale matches WHAM's.

Ten points in the log-log (top) plot of Figure \ref{hassmwham},
i.e., 1.2\% of the total, 
are off-scale in the linear-log (bottom) plot; some of these outliers
are caused by imprecise continuum-subtraction or bleed trails near bright stars,
some represent small fractions of very bright intensities, and the
remainder occur at regions with large variation in surface brightness across
the 1\arcdeg~diameter WHAM beam and may be caused by the inaccuracy in
our modeling it as a uniform circular disk even though it is slightly tapered
near its edge.

The WHAM survey is more sensitive than ours, and because of the spectral 
information it contains, the foreground  emission can be removed, providing 
absolute intensities, whereas we must resort to the method of fitting 
neighboring fields (Section \ref{georemoval}) to put all of our images on the 
same intensity scale. On the other hand, our survey has greater angular 
resolution, which is important in the study of structure at small angular 
scales, such as in filamentary nebulae. To illustrate this point, in Figure 
\ref{OriEri} we show  the Orion/Eridanus region as observed in both surveys. 

\subsubsection{The AAO/UKSchmidt Survey and the Marseille Observatory Deep 
H$\alpha $ Survey} \label{highres}

At the other end of the angular resolution scale are the photographic survey of 
the southern Galactic plane with the UK Schmidt Telescope of the Anglo-
Australian Observatory \citep{par98}, and the deep H$\alpha $ survey by the 
Marseille Observatory group with a scanning Fabry-Perot interferometer located 
at the European Southern Observatory in Chile \citep{russ98,lecor92}. The former 
will provide images of $1\arcsec $ resolution over a $5\fdg5$ field at a 
sensitivity of 5-10 R, but will be limited to the Galactic plane ({$\rm |b| < 
10\degr$}). The latter has  lower resolution ($9\arcsec $ pixels) and a smaller 
field of view ($0\fdg6 \times 0\fdg6$) but higher sensitivity (0.2 R) and of 
course gives velocity information lacking in an imaging survey. But it covers a 
very limited region of the Galactic plane ($234\degr < l < 355\degr$ and $-
2\degr < b < +2\degr $). Figure \ref{highres} compares one of our images with a 
corresponding field from the AAO/UKST survey and one of the fields from the 
Marseille survey, both of the latter taken from \citet{geor00} (Figs. 1 and 6).


\subsection{Access to the Survey Data}  \label{access}

Images of individual fields in the survey may be examined on the web at \break 
\url{http://amundsen.swarthmore.edu/SHASSA/} or 
\url{http://www.lco.org/SHASSA/}. Four images, similar to those shown in Figure 
\ref{OrionQuad},  are available for each location: H$\alpha $, continuum, 
continuum-subtracted, and smoothed (see Section \ref{dataprocessing} for a 
description of the processing involved in generating these image types). 
Individual images in FITS format may be downloaded via the above web page. Each 
full image file contains about 4 Mbytes, but compressed versions (in gzip 
format) are also available, averaging about 1 Mbyte each.

Those who wish to download a large number of images may find it more convenient 
to do so via the anonymous ftp server \url{amundsen.swarthmore.edu/SHASSA/} 
or the alternative server \url{ftp.lco.org/SHASSA/}.

All of the survey images can also be obtained (in compressed form) as a set of 
three CD-ROMs. Contact Wayne Rosing (\email{wrosing@lco.org}) for details.

Those who use the survey images in a publication or publish research based on 
the survey images should include an acknowledgement to the ``Southern H-Alpha 
Sky Survey Atlas, which was produced with support from  the National Science 
Foundation'' and make reference to this paper.

{\bf Reminder:} Several factors may affect the photometric accuracy of
the images in this atlas:

a) The removal of the geocoronal foreground emission (Section \ref{georemoval})
 is intended to offset each image so as to make them all consistent with
 sparsely-sampled WHAM data north of declination -30\arcdeg. The
 accuracy of that process is discussed in Section 6.2.2 and displayed in
Figure 13. South of
 declination -30\arcdeg, the ``background'' intensity of each image is
 unverified and for some applications should be treated as a free
 parameter.

b) [NII] emission may contribute significantly to the ``H$\alpha $''
brightness
and [SII] emission may distort the ``continuum'' brightness of some
astronomical objects 
(Sections \ref{filters} and \ref{intensityscale}).

c) Telluric OH emission may contaminate both the H$\alpha $ and
continuum images
 (Sections \ref{filters}, \ref{continuumsubtraction}, and \ref{corrections}).

d) Corrections for other instrumental effects (Sections \ref{calibrations} 
and \ref{corrections}) may distort
the relative
brightness across some objects.

Users are advised to consider carefully how the reduction procedures
described in this paper may
affect their particular photometric applications.

\subsection{Examples of Scientific Uses of the Survey}\label{examples}

We have purposely excluded new scientific analysis from this paper. However, 
papers already published illustrate some of the scientific applications for 
which the type of data in this survey is appropriate. 

First, the large size of the images ($13\degr \times 13\degr$) allows objects, 
such as the Large Magellanic Cloud, to be seen in their entirety without the 
need for piecing together a mosaic \citep{grmv99}. 

Second, objects of low surface brightness but large angular size will be found 
here which have been missed in other studies. Two examples are
the shell around the planetary nebula Abell 36 \citep[]{mbgrv01} and
the 20\arcdeg-diameter filamentary
bubble centered on $(l,b) = (275\arcdeg, 19\arcdeg)$ \citep[]{mgrv01} visible to 
the upper-left
of the Gum nebula in Figure \ref{mosaicfig}.

Third, previous studies which were mainly qualitative can now be made more 
quantitative. For example, \citet{hall99}  calibrated the Sharpless brightness 
classes of diffuse nebulae on an absolute scale of rayleighs. [Hall also 
discovered faint outer extensions of several planetary nebulae with radii many 
times the radii listed in the Catalog of Galactic Planetary Nebulae 
\citep{acker92}.]

Fourth, the correlation of H$\alpha $ emission with dust emission can be very 
informative. Dust clouds, bright at 100$\micron $, often show a ``silver 
lining'' of H$\alpha $ emission on the side facing strong ionizing flux coming 
from the UV-emitting stars in the Galactic plane (\citet{mgrv99}, Figure 7). In 
another example \citep{jenk00} a comparison of 100$\micron $ and H$\alpha $ 
emission helped delineate the three-dimensional structure of the complex region 
near the Orion Belt star $\zeta $ Ori. 

Finally, our survey has  proved useful in setting limits on Galactic emission 
which might contaminate the DASI measurements of the cosmic microwave background 
\citep{dasi01}.

We expect that many other applications will be found for the images in this 
atlas.


\section{Summary} \label{summary}

In this paper, and on the accompanying web sites 
(\url{http://amundsen.swarthmore.edu/SHASSA/} or 
\url{http://www.lco.org/SHASSA/}), we  present a digital atlas of the southern 
sky ($\delta = +15\degr $ to $-90\degr $) as seen at 656.3 nm, the H$\alpha $ 
emission line of hydrogen. The atlas consists  of 542 fields, each field 
covering a  $13\degr \times 13\degr $ region of the sky. The basic angular 
resolution is $0\farcm8 $ pixel$^{-1}$. Smoothing to a resolution of $4\farcm0 $  
allows features as faint as 0.5 rayleigh to be seen.


\acknowledgements

We are  grateful to the following students who have participated in various 
phases of this project:  Peter Austin, Gang Chen, Katherine Hall, Nini 
Khosrowshahi,  Eun Oh, Yukhi Tajima, and Anteneh Tesfaye (Swarthmore College); 
Nat Farney, Brett Schneider, and Dan Seaton (Williams College); Lynne Raschke 
and Andrew Voellmy (Haverford College); 
and Chad Bender, Ray Chen, Patrick Hentges, Dan Logan, Ian O'Dwyer, and
Jim Pulokas.(Univ. of Illinois). Gang Chen also worked  for one year as a 
practical trainee at Las Cumbres Observatory, where he was responsible for 
development of the code governing the robotic observing protocol, for the design 
and construction of the light box, and for  the initial development and testing 
of the camera system.

We are also grateful to Ron Reynolds for his advice and encouragement, and for 
providing access to some of the WHAM data in advance of publication.

The project has been supported by grants or contracts from 
the National Science Foundation (AST 9529057 \& AST 9900622 to JEG, AST-9874670 
to PRM), NASA (NAS7-1260 to JPL), a Cottrell Scholarship from the Research
Corporation to PRM, Las Cumbres Observatory, Dudley Observatory,
the Fund for Astrophysical Research,
the University of Illinois, and Swarthmore College.

The observations were obtained at Cerro Tololo Inter-American Observatory, which 
is operated by the Association of Universities for Research in Astronomy, Inc., 
under cooperative agreement with the National Science Foundation. We are 
extremely grateful to the staff of CTIO for their support of this project. 
Without their hard work on our behalf, the survey would have proceeded much more 
slowly, and might never have been completed.


%



\clearpage 

\begin{deluxetable}{lllllrr}
\tabletypesize{\scriptsize}
\tablecaption{Survey Field Centers\tablenotemark{a} \label{CentersTable}}
\tablewidth{0pt}
\tablehead{
\colhead{Field} & \colhead{RA(1950)}   & \colhead{Dec(1950)}   &
\colhead{RA(2000)} & \colhead{Dec(2000)} &
\colhead{Long.}  & \colhead{Lat.} 
}
\startdata
002 & $00^{h}00^{m}$ & -80\degr 00\arcmin  & $00^{h}03^{m}$  & -79\degr 
43\arcmin &      305 \fdg 6 & -37 \fdg 1 \\
003 & $03^{h}00^{m}$ & -80\degr 00\arcmin  & $02^{h}58^{m}$  & -79\degr 
48\arcmin &      296 \fdg 4 & -35 \fdg 7 \\
004 & $06^{h}00^{m}$ & -80\degr 00\arcmin  & $05^{h}56^{m}$  & -79\degr 
60\arcmin &      291 \fdg 8 & -29 \fdg 1 \\
005 & $09^{h}00^{m}$ & -80\degr 00\arcmin  & $08^{h}58^{m}$  & -80\degr 
12\arcmin &      294 \fdg 0 & -21 \fdg 7 \\
006 & $12^{h}00^{m}$ & -80\degr 00\arcmin  & $12^{h}03^{m}$  & -80\degr 
17\arcmin &      300 \fdg 8 & -17 \fdg 6 \\
007 & $15^{h}00^{m}$ & -80\degr 00\arcmin  & $15^{h}07^{m}$  & -80\degr 
12\arcmin &      308 \fdg 7 & -18 \fdg 9 \\
008 & $18^{h}00^{m}$ & -80\degr 00\arcmin  & $18^{h}09^{m}$  & -79\degr 
60\arcmin &      313 \fdg 8 & -24 \fdg 9 \\
009 & $21^{h}00^{m}$ & -80\degr 00\arcmin  & $21^{h}07^{m}$  & -79\degr 
48\arcmin &      313 \fdg 0 & -32 \fdg 5 \\
010 & $00^{h}00^{m}$ & -70\degr 00\arcmin  & $00^{h}03^{m}$  & -69\degr 
43\arcmin &      309 \fdg 1 & -46 \fdg 8 \\
011 & $01^{h}44^{m}$ & -70\degr 00\arcmin  & $01^{h}45^{m}$  & -69\degr 
45\arcmin &      296 \fdg 2 & -46 \fdg 7 \\
012 & $03^{h}28^{m}$ & -70\degr 00\arcmin  & $03^{h}28^{m}$  & -69\degr 
50\arcmin &      286 \fdg 0 & -41 \fdg 8 \\
013 & $05^{h}12^{m}$ & -70\degr 00\arcmin  & $05^{h}12^{m}$  & -69\degr 
57\arcmin &      280 \fdg 9 & -33 \fdg 9 \\
014 & $06^{h}56^{m}$ & -70\degr 00\arcmin  & $06^{h}56^{m}$  & -70\degr 
04\arcmin &      280 \fdg 8 & -25 \fdg 0 \\
015 & $08^{h}40^{m}$ & -70\degr 00\arcmin  & $08^{h}40^{m}$  & -70\degr 
11\arcmin &      284 \fdg 6 & -16 \fdg 9 \\
016 & $10^{h}24^{m}$ & -70\degr 00\arcmin  & $10^{h}25^{m}$  & -70\degr 
15\arcmin &      291 \fdg 1 & -10 \fdg 8 \\
017 & $12^{h}08^{m}$ & -70\degr 00\arcmin  & $12^{h}11^{m}$  & -70\degr 
17\arcmin &      299 \fdg 5 & -7 \fdg 7 \\
018 & $13^{h}52^{m}$ & -70\degr 00\arcmin  & $13^{h}56^{m}$  & -70\degr 
15\arcmin &      308 \fdg 4 & -8 \fdg 1 \\
019 & $15^{h}36^{m}$ & -70\degr 00\arcmin  & $15^{h}41^{m}$  & -70\degr 
10\arcmin &      316 \fdg 5 & -11 \fdg 9 \\
020 & $17^{h}20^{m}$ & -70\degr 00\arcmin  & $17^{h}26^{m}$  & -70\degr 
03\arcmin &      322 \fdg 5 & -18 \fdg 5 \\
021 & $19^{h}04^{m}$ & -70\degr 00\arcmin  & $19^{h}09^{m}$  & -69\degr 
55\arcmin &      325 \fdg 5 & -26 \fdg 9 \\
022 & $20^{h}48^{m}$ & -70\degr 00\arcmin  & $20^{h}53^{m}$  & -69\degr 
49\arcmin &      324 \fdg 4 & -35 \fdg 7 \\
\nodata & \nodata & \nodata & \nodata & \nodata & \nodata & \nodata \\
\enddata
\tablenotetext {a} {Table 1 is published in its entirety in the electronic 
edition of PASP,
and on the web sites \url{http://amundsen.swarthmore.edu/SHASSA/} and 
\url{http://www.lco.org/SHASSA/}.
A portion is shown here for guidance regarding its form and content.}
\end{deluxetable}
 
this table separately from the paper


\clearpage

FIGURE CAPTIONS

Figure 1---Transmission of the H$\alpha$ and Continuum filters. The positions of 
interstellar emission lines and atmospheric OH emission lines are also shown.

Figure 2---Transmission of H$\alpha$ light by the H$\alpha$ Filter as a Function 
of Angle of Incidence.

The edges of our images are $\approx 6\fdg 6$ from the center, and the corners 
are $\approx 9\fdg 3$ from the center.

Figure 3---Average brightness deviation from the mean as a function of radius 
for continuum images with $|b| > 40\degr $. The vertical bars show the range of 
this function (1 $\sigma $) for individual images.

Figure 4---Continuum, H$\alpha $,  continuum-subtracted, and  smoothed images of 
a region in Orion
(counterclockwise from upper right).

Figure 5---Brightness as a function of radius for a typical continuum-subtracted 
image before correction and after two stages of correction.

Figure 6---Comparison of fluxes of compact planetary nebulae measured by us to
fluxes reported by \citet[]{dop97}.
Our fluxes have been corrected for the two [N II] emission lines adjacent to
\halpha~(see Section \ref{intensityscale}).
The PNe represented by the solid circles were
used to determine the overall gain calibration of our images
(i.e., the number of ADUs per unit \halpha~flux), thus the mean difference
between the fluxes in ``this work'' and those of
Dopita \& Hua is zero by definition of the gain factor:
$6.8 \pm 0.3$ ADU corresponds to 1 rayleigh of \halpha~emission in 1 pixel.
Fainter PNe (which were not used) are shown as open circles. The logarithms
are base 10, and the units of the fluxes are \ergscmsec.

Figure 7---This hypothetical example of four overlapping images is used
to explain the mosaicing method in Section \ref{georemoval}.
Each image is numbered and outlined by a bold solid line. A thin solid
line outlines the ``good'' region of each image, shown as solid white;
the grey region between
the thin solid line and the bold solid line indicates the part of each image
that is flagged as ``bad'' due to edge effects. Each region of overlap
between a pair of ``good'' regions is indicated by an inscribed dashed
rectangle. Circles indicate locations and sizes of regions observed by the
1\arcdeg~diameter beam of WHAM. Data within either the dashed rectangles and/or
the circles contribute to determining each image's three free parameters,
piston, tip, and tilt (\barF~and \gradF).

Figure 8---Locations of WHAM surface brightness data are plotted on this
Hammer-Aitoff projection as
plus signs ($+$); these correspond approximately to 4\arcdeg~spacing
in right ascension and 5\arcdeg~spacings in declination from -25\arcdeg~to
+10\arcdeg.
Those WHAM points used to determine the parameters of the
fit in Equation \ref{cscblaw}
are enclosed in squares.
The dots ($\cdot$) are spaced 4\arcdeg~in right ascension and 5\arcdeg~in
declination from $-85$\arcdeg~to $-35$\arcdeg, and with Galactic latitude
$b < -20$\arcdeg, but excluding regions near the LMC,
SMC, and extended parts of the Gum nebula.
Equation \ref{cscblaw} is evaluated at the locations indicated by dots
in order to constrain the
fit for each individual image's piston, tip, and tilt in the southern sky
as described in Section \ref{georemoval}.

Figure 9---Equation \ref{cscblaw} is plotted here as a solid line and is a
fit to the subset of WHAM data defined in the previous
Figure (the plus signs enclosed by squares).
The data plotted here are from WHAM (Reynolds 2001).

Figure 10---Some interesting fields from the atlas: 
108--- l = $265\fdg3$ , b = $+19\fdg5$, filaments in Antlia;
204--- l = $190\fdg3$ , b = $-36\fdg7$, part of the Eridanus Filament;
640--- l = $255\fdg5$ , b = $+18\fdg5$, a bright rim in Pyxis;
576--- l = $261\fdg5$ , b = $-5\fdg5$, part of the Gum Nebula and the Vela SNR;
150--- l = $353\fdg5$ , b = $+22\fdg7, \delta$ Sco;
187--- l = $7\fdg3$ , b = $+22\fdg7, \zeta $ Oph.

Figure 11---Hammer-Aitoff projection in Galactic coordinates
of our continuum-subtracted \halpha~data.

Figure 12---Comparison of our continuum-subtracted \halpha~images to those of 
the
VTSS survey shows them to be similar with a few differences (see
Section \ref{others}~for details not in this caption). Our data (left
image above), rendered to match one of the VTSS images (right image above),
illustrates the comparison. 
The VTSS image, Aql04, 10\arcdeg~in diameter, centered on 
$19^h08^m~+5$\arcdeg~[2000], with North up and East to the left,
was selected at random for the comparison, and from it an arbitrary 2-D plane of
emission was subtracted in order to minimize the residuals between it and the
image rendered from our data.
The lower-left plot shows a pixel-by-pixel
comparison of the data from those two images after convolution with a
$5\times 5$ pixel ($8\arcmin \times 8\arcmin$) kernel.
If the intensity calibration
of our data and the VTSS data were the same, the data would scatter
about the diagonal solid line. The dashed line is a first-order fit (with
zero intercept) to the data with slope 0.85.
A similar plot made for the VTSS region Ori11, centered on
$5^h24^m~-9$\arcdeg~[2000], is shown at lower-right. It has
a large enough dynamic range of surface brightness to allow a good comparison
of the VTSS intensity calibration with our own. The slope of the
dashed line in the lower-right plot is 0.80.

Figure 13---The \halpha~surface brightness measured by WHAM and by this work
are compared. 
This work's values are the average of our data within the
1\arcdeg-diameter WHAM beam.
In the top plot, the diagonal line indicates equality of the two
surface brightness measurements and the two additional lines are
${\rm Y = (1 \pm 0.1) X \pm 0.4}$ R,
where X is the abscissa and Y is the ordinate
of the plot. The bottom plot shows the same data as the top plot,
but in the form of residuals (our values minus WHAM's) compared to
WHAM's values; in the bottom plot, the two curved lines are
${\rm Y = \pm (0.8 + 0.1 X)}$ R. The r.m.s. of the residuals
of points with WHAM-measured surface brightness less than 2 R is 0.4 R.

Figure 14---Our continuum-subtracted \halpha~data (upper image) and the WHAM
survey data (lower image) of the same region, Orion/Eridanus,
illustrate the effect of improved angular resolution in filamentary
regions. The upper image has 2\arcmin~resolution; the lower image
has 60\arcmin~resolution and has been interpolated between WHAM beams
(data provided by \citet[]{haff01}).

Figure 15---Our field 033 (top), compared to field HA176 of the AAO/UKST survey 
(left) and the region around Gum 37 from the Marseille Observatory survey 
(right). The second image is scanned from  \citet{geor00} and  does not 
represent the full image quality of the AAO/UKST survey.

\end{document}